\documentclass[sigconf]{acmart}

\AtBeginDocument{%
  }

\setcopyright{acmlicensed}
\copyrightyear{2025}
\acmYear{2025}

\acmConference[CONSEQUENCES ‘25 workshop, co-located with ACM RecSys ‘25]{October,
  2025}

\begin{document}

\title{Algorithm Adaptation Bias in Recommendation System Online Experiments}


\author{Chen Zheng}
\email{czheng@roblox.com}
\affiliation{%
  \institution{Roblox}
  \city{San Mateo}
  \state{California}
  \country{USA}
}

\author{Zhenyu Zhao}
\email{zzhao@roblox.com}
\affiliation{%
  \institution{Roblox}
  \city{San Mateo}
  \state{California}
  \country{USA}
}



\begin{abstract}
Online experiments (A/B tests) are widely regarded as the gold standard for evaluating recommender system variants and guiding launch decisions. However, a variety of biases can distort the results of the experiment and mislead decision-making. An underexplored but critical bias is \emph{algorithm adaptation effect}. This bias arises from the flywheel dynamics among production models, user data, and training pipelines: new models are evaluated on user data whose distributions are shaped by the incumbent system or tested only in a small treatment group. As a result, the measured effect of a new product change in modeling and user experience in this constrained experimental setting can diverge substantially from its true impact in full deployment. In practice, the experiment results often favor the production variant with large traffic while underestimating the performance of the test variant with small traffic, which leads to missing opportunities to launch a true winning arm or underestimating the impact. 
This paper aims to raise awareness of algorithm adaptation bias, situate it within the broader landscape of RecSys evaluation biases, and motivate discussion of solutions that span experiment design, measurement, and adjustment. We detail the mechanisms of this bias, present empirical evidence from real-world experiments, and discuss potential methods for a more robust online evaluation.
\end{abstract}


\maketitle

\section{Introduction}

Recommender systems (RecSys) are central to user experiences on major digital platforms, determining the content, products, or connections users encounter~\cite{knightcolumbia2023}. Online experiments, particularly A/B tests, are the standard approach for evaluating algorithmic changes, aiming to estimate causal effects on engagement and business metrics through randomized treatment assignment. This setup is based on the stable unit treatment value assumption (SUTVA)~\cite{imbens2015causal}, which assumes that there is no interference between the units and a consistent treatment application.

Although prior work has explored SUTVA violations in networked or adaptive settings~\cite{zhu2024treatment, karrer2021network, saint2019using, kohavi2020trustworthy}, a subtle yet pervasive source of bias remains underexplored: \emph{algorithm adaptation effect}.

In typical experiments, a small user segment receives the new model, while the majority remains under the incumbent system. However, user behavior, content popularity, and creator incentives are shaped by the production model, forming a feedback loop that also informs training data for future models. As a result, treatment variants can underperform in limited-rollout experiments, not due to inferior quality, but because they fail to trigger dynamics such as content virality, creator adaptation, or sufficient data accumulation.

This misalignment causes bias: the measured treatment effect no longer reflects the true causal effect of full deployment. The bias often favors the dominant-traffic variant (typically the control), leading to understated impact or missed opportunities for beneficial launches.

Examples of affected experiments include: (1) algorithmic changes in retrieval or ranking (e.g., adding high-value actions to ranking objectives) and (2) UI modifications that influence content presentation (e.g., thumbnails or metadata).

This paper makes the following contributions: 
\begin{itemize}
    \item Formalizes the algorithm adaptation effect within the broader taxonomy of RecSys evaluation biases.
    \item Presents empirical evidence from real-world experiments.
    \item Proposes experimental and analytical techniques for more robust measurement.
    \item Outlines future directions for better detection and mitigation.
\end{itemize}

\section{Background and Related Work}

Our focus is on \emph{algorithm adaptation bias} in measuring treatment effects from online experiments. While related, this is distinct from the broader literature on algorithmic bias in recommender systems, which examines structural sources of bias such as popularity, selection, exposure, and position bias~\cite{boratto2020hands, stinson2022algorithms, sciencedirect2022}. These biases can degrade recommendation quality, reinforce inequality, and lead to misalignment between offline and online performance~\cite{redgate2023}.

\begin{itemize}
    \item \textbf{Popularity Bias}: Tendency to overrecommend popular items, reducing visibility for niche or long-tail content~\cite{sciencedirect2022}.
    \item \textbf{Selection Bias}: Recommendations shape the observed feedback, creating self-reinforcing loops~\cite{stinson2022algorithms}.
    \item \textbf{Exposure Bias}: Unshown items receive little or no feedback, limiting learning and visibility.
    \item \textbf{Position/Presentation Bias}: UI affects user attention and interaction, such as higher-ranked items get more clicks, regardless of relevance.
\end{itemize}

Although algorithmic bias is well studied, the challenge of \emph{algorithm adaptation bias} - where experimental outcomes are distorted by the embedded feedback loops of the production system - remains underexplored.

A more adjacent area is experimentation bias in networked or interactive environments, particularly on social platforms~\cite{zhu2024treatment, karrer2021network, saint2019using, kohavi2020trustworthy}. However, such work rarely addresses the bias introduced by adaptive recommender systems specifically.

\section{Algorithm Adaptation Bias in Online Experiments}
\label{sec:alg-adapt-bias}

\subsection{Formulation}

We study the setting in which a platform currently implements a \emph{production} recommendation policy $\pi_0$ and evaluates a \emph{candidate} policy $\pi_1$ through an online A / B test with a (typically small) share of treatment $\rho \in (0,1)$. Let $Y$ denote a primary outcome of interest (e.g., revenue, retention, click through rate or play through rate, deep engagement). Because recommendation policies shape the data distribution they are trained and evaluated on (via user behavior, creator incentives, content supply, etc.), deploying $\pi_1$ only on a small fraction of traffic does \emph{not} induce the same user--item distribution that would arise if $\pi_1$ were fully launched. Consequently, the standard difference-in-means estimator in the experiment targets an estimand that can differ systematically from the platform’s \emph{policy-level causal effect} under a full rollout.

\paragraph{Target (policy-level) estimand.}
Let $\mathcal{D}(\pi)$ denote the stationary / equilibrium distribution of users, items, and contexts that emerges when policy $\pi$ is deployed platform-wide.\footnote{Formally, one can view $\mathcal{D}(\pi)$ as the distribution induced by the data-generating process (users, creators, ranking, feedback, re-training, etc.) under the stable deployment of $\pi$; see also performative prediction \citep{perdomo2020performative} and causal interference frameworks \citep{hudgens2008toward, imbens2015causal, eckles2017design, karrer2021network}.} The \emph{policy-level ATE} we care about is
\begin{equation}
\tau^\star \;=\; \mathbb{E}_{\mathcal{D}(\pi_1)}[Y(\pi_1)] - \mathbb{E}_{\mathcal{D}(\pi_0)}[Y(\pi_0)].
\label{eq:policy-ate}
\end{equation}
This is the effect we would obtain by replacing $\pi_0$ with $\pi_1$ for the entire population (and letting the system re-equilibrate).

\paragraph{Experimental estimand under a partial rollout.}
In the online experiment, only a fraction $\rho$ of the population is exposed to $\pi_1$, while the remaining $1-\rho$ continues to interact with $\pi_0$. Let $\Pi_\rho$ denote the \emph{mixture deployment} (control on $1-\rho$, treatment on $\rho$). Let $\mathcal{D}_\rho$ be the induced distribution of users, items, and contexts under $\Pi_\rho$ (this generally differs from both $\mathcal{D}(\pi_0)$ and $\mathcal{D}(\pi_1)$ because users, items, and creators interact across variants). The standard difference-in-means estimand is
\begin{equation}
\tau_{\text{exp}}(\rho) \;=\; 
\mathbb{E}_{\mathcal{D}_\rho}\!\left[\, Y(\pi_1) \mid Z=1 \,\right]
-
\mathbb{E}_{\mathcal{D}_\rho}\!\left[\, Y(\pi_0) \mid Z=0 \,\right],
\label{eq:exp-estimand}
\end{equation}
where $Z \in \{0,1\}$ is the treatment assignment. When there is \emph{no} interference or feedback-loop adaptation, $\mathcal{D}_\rho$ factors and $\tau_{\text{exp}}(\rho)$ coincides with $\tau^\star$. In RecSys, however, both user behavior and supply-side responses are shaped by the dominant policy, violating SUTVA \citep{imbens2015causal} and inducing interference \citep{hudgens2008toward,karrer2021network,eckles2017design}.

\begin{definition}[Algorithm Adaptation Bias]
The \emph{algorithm adaptation bias} at treatment share $\rho$ is the gap between the experimental estimand and the full-deployment policy estimand:
\begin{equation}
\mathsf{Bias}(\rho) 
\;=\; 
\tau_{\text{exp}}(\rho) - \tau^\star.
\label{eq:bias-def}
\end{equation}
\end{definition}

Intuitively, $\mathsf{Bias}(\rho)$ captures how much we mis-estimate the platform-wide causal effect of launching $\pi_1$ when we measure it in a world still largely governed by $\pi_0$.

\paragraph{A simple decomposition.}
Adding and subtracting $\mathbb{E}_{\mathcal{D}_\rho}[Y(\pi_1)]$ and $\mathbb{E}_{\mathcal{D}_\rho}[Y(\pi_0)]$ yields
\begin{align}
\mathsf{Bias}(\rho)
&= 
\underbrace{
\left(
\mathbb{E}_{\mathcal{D}_\rho}[Y(\pi_1)] - \mathbb{E}_{\mathcal{D}(\pi_1)}[Y(\pi_1)]
\right)}_{\text{adaptation gap for }\pi_1}
-
\underbrace{
\left(
\mathbb{E}_{\mathcal{D}_\rho}[Y(\pi_0)] - \mathbb{E}_{\mathcal{D}(\pi_0)}[Y(\pi_0)]
\right)}_{\text{adaptation gap for }\pi_0}.
\label{eq:bias-decomp}
\end{align}
The first term reflects that the treatment arm is evaluated in a distribution \emph{dominated by} the control policy (when $\rho$ is small), preventing $\pi_1$ from realizing ecosystem-level gains (e.g., creator adaptation, social spillovers, cold-start relief). The second term acknowledges that the control arm is also evaluated off its full-deployment distribution, although in practice $\mathcal{D}_\rho$ is typically much closer to $\mathcal{D}(\pi_0)$ when $\rho \ll 1$, which often makes the overall bias negative (i.e., systematically favoring the production policy).

The feedback loop between algorithms and the data they induce closely relates to \emph{performative prediction} \citep{perdomo2020performative} and to concerns about adaptive experiments and biased estimators in bandit/RL settings \citep{hadad2021confidence}. We differ by centering the bias on the RecSys \emph{launch decision} problem: the estimand of interest is the \emph{full-deployment} effect, while the experiment is run in a \emph{mixture} world that typically advantages the incumbent.

\subsection{Mechanisms and Causes}
Algorithm adaptation bias arises from the dynamic interplay between recommendation algorithms and the ecosystem they operate in. When the test policy $\pi_1$ is deployed to only a small subset of users, its ecosystem-level impact is constrained in several ways. We outline key mechanisms that contribute to the resulting bias:

\begin{itemize}
    \item \textbf{Insufficient Ecosystem Amplification:} Content retrieved or promoted by the treatment model may fail to gain traction due to insufficient exposure volume. Niche or novel items that require network effects (e.g., trending, virality, social proof) may underperform in limited rollout.
    
    \item \textbf{Social and Multiplayer Dynamics:} In domains with co-play or social engagement (e.g., multiplayer games, social feeds), the treatment group may lack sufficient concurrent users to support the intended experience, leading to a systematic underestimation of the treatment effect.
    
    \item \textbf{Feedback-Driven Training Pipelines:} The training data used by both control and treatment models is typically influenced by the dominant production policy. This can bias evaluations in favor of the incumbent, as the treatment model does not benefit from feedback on content or users it would have influenced under full deployment.
    
    \item \textbf{Creator and Supply-Side Response Lag:} Content creators and producers often adapt their behavior (e.g., style, format, upload timing) in response to new ranking signals. These adaptations are unlikely to occur unless the new model is sufficiently deployed to generate observable effects, reducing the measured gains of the test policy.
    
    \item \textbf{Exposure-Incentive Misalignment:} Incentive structures (e.g., monetization, rewards) may remain aligned with the control policy during the test phase, deterring optimal user and creator behavior under the treatment model.
\end{itemize}

These mechanisms highlight that the algorithm adaptation bias is not merely a statistical artifact, but a structural consequence of deploying new models in a system governed by legacy dynamics. As such, they motivate the need for experimental designs and estimators that explicitly account for these interactions.

\section{Empirical Evidence from Online Experiments}

We present two example cases where online experiments exhibited signs of algorithm adaptation bias. These cases motivate the need for systematic evaluation methods tailored to detect such bias.

\subsection{UI Experiment: Thumbnail Format Redesign}

A redesigned content thumbnail format in the recommendation feed was launched in two phases across different user segments. In both phases, pre-launch A/B tests showed neutral or non-significant impact on key metrics. However, post-launch analyses consistently revealed a positive lift. This divergence suggests the presence of adaptation bias favoring the production variant during pre-launch.

Two plausible mechanisms may explain the shift: 1) \textbf{Creator Adaptation:} After the feature's initial release, creators updated assets (e.g., thumbnails) to better align with the new format, improving effectiveness post-launch. 2) \textbf{Algorithmic Feedback:} Content favored by the new UI received more exposure and engagement, triggering feedback loops that amplified its presence and reinforced performance gains.

\subsection{Algorithm Experiments}

Similar patterns emerged in algorithm-focused experiments. Changes to ranking objective terms or loss function weights showed small positive lifts in pre-launch experiments, which were consistently higher in post-launch evaluations. These results further illustrate how adaptive system dynamics can distort effect estimates during partial rollouts.

These findings provide early empirical support for algorithm adaptation bias and highlight the need for evaluation frameworks that account for ecosystem-level feedback and model-data co-adaptation.

\section{Measurement and Mitigation Methods}

We outline several promising methods to quantify and mitigate algorithm adaptation bias.

\subsection{Model-Data Separation}

A principled approach is to decouple model training data in experiments from production traffic. Each variant is trained solely on data from its corresponding experiment group, with balanced traffic allocation. While this design ensures fairer comparisons, it can be prohibitively expensive to implement at scale due to infrastructure complexity.

\subsection{Adaptive Experiment Design with Ramp-Up}

Staging the rollout of the treatment variant—particularly including a 50/50 traffic split phase—can help surface adaptation effects. By comparing estimated effects at different traffic levels, practitioners can observe whether the treatment impact grows as it gains broader exposure. While not perfectly unbiased, this method offers a closer approximation to full-deployment effects.

\subsection{Confirmation Analysis}

To detect adaptation bias and uncover underlying mechanisms, we recommend a set of post-hoc diagnostics: 1) \textbf{Impression Shift Analysis:} Compare content-level exposure before and after ramp-up to identify items benefiting from ecosystem adaptation; 2) \textbf{Engagement Trajectories:} Track user satisfaction metrics for content favored by the treatment policy to detect amplification effects at scale; 3) \textbf{Popularity Distribution Shift:} Examine how content impression distributions evolve, assessing whether the treatment increases diversity or reduces concentration on already-popular items.

These methods offer practical tools for surfacing and understanding adaptation-driven discrepancies between pre-launch and post-launch evaluation.

\section{Conclusion}

This paper aims to raise awareness and stimulate further exploration of \emph{algorithm adaptation bias} in the evaluation of online experiments for recommender systems. While we outline the potential severity of this bias and provide initial formalization and empirical evidence, additional research is required to deepen understanding and develop practical mitigation strategies, especially on forming bias estimation and adjustment methods.

\bibliographystyle{ACM-Reference-Format}
\bibliography{reference}

\appendix
\section{Appendix}

\end{document}